\documentclass[12pt]{article}
\usepackage{epsf}
\def\be{\begin{equation}}
\def\ee{\end{equation}}
\def\bea{\begin{eqnarray}}
\def\eea{\end{eqnarray}}

\begin{document}
\begin{titlepage}
\begin{center}
{\Large \bf William I. Fine Theoretical Physics Institute \\
University of Minnesota \\}
\end{center}
\vspace{0.2in}
\begin{flushright}
FTPI-MINN-13/13 \\
UMN-TH-3201/13 \\
April 2013 \\
\end{flushright}
\vspace{0.3in}
\begin{center}
{\Large \bf $Z_c(3900)$ -- what is inside?
\\}
\vspace{0.2in}
{\bf M.B. Voloshin  \\ }
William I. Fine Theoretical Physics Institute, University of
Minnesota,\\ Minneapolis, MN 55455, USA \\
School of Physics and Astronomy, University of Minnesota, Minneapolis, MN 55455, USA \\ and \\
Institute of Theoretical and Experimental Physics, Moscow, 117218, Russia
\\[0.2in]

\end{center}

\vspace{0.2in}

\begin{abstract}
The models for the internal structure of the newly found four-quark charmonium-like resonance $Z_c(3900)$ are discussed: the molecular model as well as the hadro-charmonium and tetraquark schemes. It is argued that it would be possible to resolve between these models by combining measurements of the quantum numbers of the resonance and of its decay rates into yet unseen channels $\pi \psi'$, $\pi h_c$, $\rho \eta_c$ and into pairs of heavy mesons $D^* \bar D$ and $D \bar D^*$. The models also predict different related four-quark states, which can be sought for in the existing and future data.
\end{abstract}
\end{titlepage}

The reported by BESIII~\cite{beszc} charged charmonium-like peak $Z_c(3900)$ in the channel $\pi^\pm J/\psi$ is the newest addition to the `collection' of known states related to heavy quarkonium that manifestly require the presence of two quarks and two antiquarks in their composition. Other hadrons of this type are the peaks in the $\pi^\pm \psi'$ and $\pi^\pm \chi_{c1}$ spectra reported by Belle~\cite{bellez4,bellez12} (although not confirmed by $BABAR$ searches~\cite{babarz4,babarz12}) and the bottomonium-like `twin' resonances $Z_b(10610)$ and $Z_b(10650)$ produced in the decays $\Upsilon(5S) \to \pi^\pm Z^\mp_b$ and observed as peaks in the invariant mass spectra in the channels with bottomonium~\cite{bellezb} ($\pi^\pm \Upsilon(nS)$ with $n=1,2,3$ and $\pi^\pm h_b(kP)$ with $k=1,2$) and with pairs of $B$ ($B^*)$ mesons~\cite{bellebb}. Clearly, such states present a challenge for theoretical description of four-quark systems, and a better understanding of the internal workings of these and similar, yet unobserved, resonances may provide new insights into the strong dynamics of multiquark systems.

The masses and the electric charge of the discussed resonances definitely require the quark composition to be $Z_Q^+ \sim Q \bar Q u \bar d$ with $Q$ standing for $c$ or $b$, and the models discussed in the literature differ in the picture of clustering the quarks and antiquarks in this four-quark system that is used to somewhat organize and simplify the description. The so-far discussed models can be classified as follows.

\noindent
{\it Hadronic molecules}\cite{ov}. Heavy-light quark-antiquark pairs form heavy mesons, and the meson-antimeson pair moves at distances longer than the typical size of the meson. The mesons are interacting through exchange of light quarks and gluons, similar to nuclear force.

\noindent
{\it Hadro-quarkonium}\cite{mv,dv}. The $Q \bar Q$ pair forms a tightly bound system whose wave function is close to that of one of the heavy quarkonium states. The heavy quark pair is embedded in a spatially large excited state of light mesonic matter and interacts with it by a QCD analog of Van der Waals force.

\noindent
{\it Tetraquarks}\cite{mppr}. The pairs $Qq$ and $\bar Q \bar q$ form relatively tightly bound diquark and antidiquark, which interact by the gluonic color force.

Clearly, other four-quark configurations are logically possible, e.g. a more uniform state where no significant pairing occurs. Furthermore, most likely all types of configuration consistent with the overall quantum numbers are, to some extent, present in the wave function and are quantum-mechanically mixed, and the difference between the discussed clustering models is in the assumed prevalent configuration with the other ones being considered as a relatively small admixture. Given this approximate classification, it is quite likely that the observed (and yet unobserved) four-quark states exhibit different type of the dominant behavior~\cite{mv}. For instance, a description~\cite{bgmmv} of the $Z_b(10610)$ and $Z_b(10650)$ resonances as being (dominantly) molecules made of respectively $B\bar B^*$ ($B^* \bar B$) and $B^* \bar B^*$ pairs agrees with the relative strength and phase of the coupling of these particles to para- and ortho- bottomonium. On the other hand, the `affinity', of some of the charmonium-like four-quark states to a particular state of charmonium (e.g. 
$Z(4.43) \to \pi \psi'$, $Z_{1,2} \to \pi \chi_{c1}$) indicates that they likely contain that particular state embedded in the light-matter excitation in the dominant part of the wave function.

Currently it is not yet clear where the newly discovered resonance $Z_c(3900)$ fits in this classification, and an interpretation and further studies of this four-quark state can be quite instrumental in gaining understanding of multiquark heavy-light systems. The so far discussed models describing this new peak include a $D \bar D^*$ molecule or a cusp in the $D \bar D^*$ spectrum~\cite{whz} as well a molecular or tetraquark structure~\cite{fmpppr}. It has been concluded~\cite{whz} that a molecular picture is likely preferred over the cusp hypothesis on the basis of the observed shape of the $Z_c(3900)$ peak, while in Ref.~\cite{fmpppr} detailed predictions of the tetraquark model for this resonance as well as expectation for related states are discussed and some of similar properties within the molecular model are also mentioned. In this paper I concentrate on the behavior that should be expected in the molecular and hadro-charmonium models of the new state, and the ways of distinguishing by further experimental studies between the still open possibilities for its interpretation within the models. It will be argued that different models give distinctively different expected patterns of relative rates for the yet unobserved decays of the $Z_c(3900)$ resonance to the final states $\pi \psi'$, $\pi h_c$, $\rho \eta_c$ and $D \bar D^*$ as well as different predictions for other related resonances. In what follows I first discuss the expected properties specific to the molecular model and then to the hadro-charmonium picture.

In the molecular model the $Z_c(3900)$ is viewed as a resonance made of $D \bar D^*$ and $D^* \bar D$ pairs in the isovector state with positive $G$ parity. (The $I^G = 1^+$ assignment directly follows from the discovery mechanism for the resonance production: $Y(4260) \to \pi Z_c(3900)$.) If the heavy meson pair is in the  $S$ wave, as is also assumed in Refs.~\cite{whz,fmpppr}, the spin-parity of the resonance is uniquely determined as $J^P=1^+$. In this case the pion in the decay $Y(4260) \to \pi Z_c(3900)$ is emitted in the $S$ wave\footnote{The original paper \cite{beszc} mentions a fit of the $Z_c$ resonance peak under the assumption that the pion is emitted in the $P$-wave. It is however not clear whether there is an indication in the data that this process is a $P$-wave one, or that the assumption is {\it ad hoc}. Clearly, if the experiment would point toward a $P$-wave emission, the parity of the $Z_c$ resonance would have to be negative, and any discussion of it as an $S$-wave $D \bar D^*$ molecule would be totally irrelevant.}, and the chiral symmetry requires the amplitude of this process to be proportional to the pion energy $E_\pi$ (similarly to the behavior in the decays $\Upsilon(5S) \to \pi Z_b$~\cite{bgmmv}):
\be
A(Y \to \pi Z_c)  \propto E_\pi \, (\vec Y \cdot \vec Z)~
\label{api}
\ee
with $\vec Y$ and $\vec Z$ being the polarization amplitudes for $Y$ and $Z_c$. In this picture the $Z_c(3900)$ resonance is a direct charmonium-like analog of the bottomonium-like $Z_b(10610)$ resonance. Then a natural question arises of where is the analog of the higher $Z_b(10650)$ state? This leads to the expectation~\cite{fmpppr} that in this model there should be a similar `twin' resonance $Z_c'$ with the mass positioned relatively to the $D^* \bar D^*$ threshold similarly to positioning of the $Z_c(3900)$ relatively to the $D \bar D^*$ threshold. The measured mass of the $Z_c(3900)$ is $M(Z_c)=(3899.0 \pm 3.6 \pm 4.9)\,$MeV, so that the central value is 23.7\,MeV above the $D^{*+} \bar D^0$ threshold and 22.2\,MeV above that for $D^+ \bar D^{*0}$. Placing the  $Z_c'$ resonance by the same amount above the $D^{*+} \bar D^{*0}$ threshold gives its expected mass at approximately 4030\,MeV.  Assuming, as is the case for the $Z_b(10610)$ and $Z_b(10650)$ resonances, that the proportionality coefficients in the amplitudes, given by Eq.(\ref{api}), are approximately the same for the $Z_c$ and $Z_c'$, one can estimate
\be
{\Gamma[Y(4260) \to \pi Z_c']\over \Gamma[Y(4260) \to \pi Z_c]} \approx 0.22~, ~~~{\rm and} {\Gamma(Z_c' \to pi J/\psi) \over 
 \Gamma(Z_c \to pi J/\psi)} \approx 1.6~,
\label{rg}
\ee
so that the ratio of the combined transition rates through the two charmonium-like resonances can be estimated as
\be
{Br[Y(4260) \to \pi^\pm Z_c^\mp (4030) \to \pi^+ \pi^- J/\psi] \over Br[Y(4260) \to \pi^\pm Z_c^\mp (3900) \to \pi^+ \pi^- J/\psi]} \approx 0.35~.
\label{rb}
\ee
No peak of such significance is apparent in the data presented in Ref.~\cite{beszc}. However it is quite important that a dedicated experimental study of the presence of a peak near the mass 4030\,MeV similar to the $Z_c(3900)$ be done and an upper limit on its significance  established. It is in principle possible that the ratio of the combined transition rates is somewhat smaller than the estimate in Eq.(\ref{rb}), e.g. due to a lager total width of the higher $Z_c(4030)$ resonance, which can be due to its coupling to the $D^* \bar D$ channel. This coupling, which is suppressed by the heavy quark spin symmetry~\cite{bgmmv}, can be enhanced for charmonium-like states in comparison with the behavior of the $Z_b$ resonances due to lighter mass of the charmed quark.  

The $S$-wave $D^* \bar D$ molecular interpretation of the $Z_c(3900)$ resonance also implies distinctive properties of this state with regards to the total spin $S$ of the $c \bar c$ quark pair. Namely, within this interpretation the spins of the heavy quark and antiquark are not correlated with each other, but rather with the corresponding light antiquark and quark. As a result, in the molecular state the spin state of the $c \bar c$ pair is a mixture of $S=0$ and $S=1$. Specifically, in the $I^G(J^P)=1^+(1^+)$ meson pair these two components are mixed with equal weight~\cite{bgmmv}. This behavior is well known for the $Z_b$ resonances, which couple with approximately equal strength to channels with ortho-bottomonium ($\pi \Upsilon(nS)$) and with para-bottomonium ($\pi h_b(kP)$). Clearly, the same behavior should be expected of a molecular $Z_c(3900)$, i.e. in addition to its decay into $\pi J/\psi$ (and $\pi \psi'$ discussed further in this text), it should have a comparable rate of decay into final states with para-charmonium: $Z_c \to \pi h_c$ and $Z_c \to \rho \eta_c$. It should be noted that in this regard the molecular picture is somewhat similar to the tetraquark model, where the spin-correlated pairs are $c q$ and $\bar c \bar q$, so that the state of the total spin of the $c \bar c$ pair is mixed. The ratio of the decay rates in the tetraquark
model~\cite{fmpppr} is estimated as 
\be
{\Gamma(Z_c^+ \to \rho^+ \eta_c) \over \Gamma(Z_c^+ \to \pi^+ J/\psi)} \approx 0.65~,
\label{retac}
\ee
which estimate does not look unreasonable in the molecular model as well.

The known behavior of the $Z_b$ resonances is that their pion transitions to excited $\Upsilon(2S)$ and $\Upsilon(3S)$ states are not suppressed (and rather enhanced) as compared to the transition to the lowest $\Upsilon(1S)$ bottomonium in spite of a significant kinematical enhancement of the latter one. This behavior is understood~\cite{lv} in terms of larger overlap of the wave function of the $b \bar b$ quark pair in a spatially large molecule with spatially larger excited bottomonium states. 
The calculations~\cite{lv} based on modeling the  wave functions of heavy quarkonium using the Cornell potential~\cite{cornell} are in a reasonable agreement with the data~\cite{bellebb} on the relative rates of the pion transitions from $Z_b$ resonances to various excitations of bottomonium. An application of the same approach to the pion transitions from $Z_c(3900)$ yields
\be
{\Gamma[Z_c(3900) \to \pi \psi'] \over \Gamma[Z_c(3900) \to \pi J/\psi]} \approx 0.4~.
\label{zcpsi2}
\ee
A similar, although a somewhat smaller estimate (about 0.3) for this ratio, is found in Ref.~\cite{fmpppr} within the tetraquark model.

It can be argued that the interpretation of $Z_c(3900)$ as an $S$-wave $D^* \bar D$ molecule possibly runs into difficulty related to its relatively high excitation energy, $\Delta \approx 23\,$MeV, over the threshold. Indeed, at such energy the characteristic momentum of the heavy mesons is $p \sim \sqrt{M_D \Delta} \approx 200\,$MeV, where $M_D$ is the mass of either of the mesons. Such momentum corresponds to a typical distance 1\,fm between the mesons, which is uncomfortably close to the generally estimated range, where the mesons start to overlap, and can not be considered as individual particles. Also the interaction between the mesons should be quite contrived in order to explain a barrier that can hold an $S$ wave resonance at about 23\,MeV above the threshold.

It is more natural if a resonance appears in a state with nonzero orbital momentum and is `held together' by the centrifugal barrier. The lowest nonzero orbital momentum corresponds to a $P$-wave motion of the heavy meson pair, which would imply negative parity for $Z_c(3900)$. In this case the options for $J^P$ are $0^-,\,1^-$ and $2^-$, and the $1^-$ assumption can likely be discarded, since such state would have a large width due to decays into pairs of pseudoscalar $D$ mesons, $D \bar D$. The $0^-$ and $2^-$ cases are interesting in that in the $I^G(J^P)=1^+(0^-)$ or $I^G(J^P)=1^+(2^-)$ state of $D^* \bar D$ and $D \bar D^*$ pairs the total spin of the $c \bar c$ pair is fixed, $S=1$. Indeed, the only combination of spin state of the heavy quark pair $S_H$ and of the angular momentum of the rest (light) degrees of freedom $J_L$ that has these quantum numbers is $1^-_H \otimes 1^+_L$ for $J^P=0^-$ and $(1^-_H \otimes 1^+_L) \, \oplus \, (1^-_H \otimes 2^+_L)$ for $J^P=2^-$. Thus if a future angular analysis finds the  $Z_c(3900)$  resonance to be a $J^P=0^-$ or a $J^P=2^-$ state, one should expect in the molecular picture that the transitions from this resonance to states of para-charmonium, $\pi h_c$ and $\rho \eta_c$, are suppressed by the heavy quark spin symmetry relative to the transitions to ortho-charmonium, $\pi J/\psi$ and $\pi \psi'$. The estimate in Eq.(\ref{zcpsi2}) for the relative strength of the latter two transitions should be applicable in this case as well. 

As is already mentioned, another possible interpretation of the newly found $Z_c(3900)$ is that it is dominantly a hadro-charmonium state, i.e. a tightly bound $J/\psi$ state of $c \bar c$ embedded in a light-quark excitated state. The observed decay $Z_c \to \pi J/\psi$ is then a de-excitation of the light-quark matter. In this picture the $Z_c(3900)$ is tantalizingly similar to the resonance $Z(4.43)$, which decays into $\pi \psi'$, and has a very similar total width of about 45\,MeV. One can then view the latter resonance as a radial excitation of the $c \bar c$ pair over the $Z_c(3900)$ in essentially the same way as $\psi'$ is the radial excitation over the $J/\psi$. The mass difference between $Z(4.43)$ and $Z_c(3900)$ is approximately 535\,MeV, which is by about 55\,MeV lower than the mass difference between $\psi'$ and $J/\psi$, and one can speculate that this difference in the excitation energy can be attributed to the difference in the interaction with the light-quark `environment' due to a larger spatial size of $\psi'$. 

In the hadro-charmonium model the resonance $Z_c(3900)$ contains the $c \bar c$ pair in a pure $S=1$ state, so that the transitions to para-charmonium, $Z_c \to \pi h_c$ and $Z_c \to \rho \eta_c$ are expected to be suppressed. Furthermore, in as much as the $c \bar c$ pair has the wave function of $J/\psi$ (with possible slight distortions due to the interaction~\cite{mv,dv}) the transition to $\psi'$, $Z_c \to \pi \psi'$, should be suppressed in comparison with the estimate in Eq.(\ref{zcpsi2}). Another expected feature of the $Z_c(3900)$ resonance viewed as hadro-charmonium is that its decay into open charm, $D^* \bar D$ and $D \bar D^*$, should be suppressed relative to the molecular case. In this respect the hadro-charmonium model is similar to the tetraquark scheme, where this rate is estimated~\cite{fmpppr} as $\Gamma(Z_c \to D^{*+} \bar D^0, D^+ \bar D^{*0}) \approx 4\,$MeV, which accounts for only a small fraction of the total width of $Z_c(3900)$. In the molecular picture the significance of these decays can be gauged by the behavior of the $Z_b$ resonances, for which the `dissociation' into heavy meson pairs constitutes $(70 \div 80)$\% of the total width~\cite{bellebb}, corresponding to the absolute rate of about 10\,MeV or larger. One can expect that the decay into open flavor mesons should be enhanced for a molecular $Z_c(3900)$ due to its higher excitation energy above the threshold, so that these channels should account for a large, if not the major fraction of the total width.

The simplest assumption about the quantum numbers of the $Z_c(3900)$ as hadro-charmonium is that it is the $J/\psi$ embedded in an $S$-wave in a spinless excitation of the light-quark matter with the quantum numbers of a pion, i.e. $J^P=0^-$, so that the overall quantum numbers of the $Z_c(3900)$ are $I^G(J^P)=1^+(1^+)$. An assumption of a nonzero spin of the light-quark excitation and/or an orbital motion of the embedded $J/\psi$ would lead to a conclusion that there should also exist two or more states with nearby masses corresponding to a `fine structure' due to the interaction of the spin of $J/\psi$ with the angular variables of the `environment'. Since there appears to be no such structure in the data~\cite{beszc}, it is reasonable to assume that the simplest arrangement of the hadro-charmonium embedding is realized in $Z_c$.

A distinctive prediction, stemming from a hadro-charmonium interpretation of $Z_c(3900)$, is that of an isovector four-quark resonance $W_c$, where the embedded $J/\psi$ is replaced with the $\eta_c$. In the limit of heavy quark spin symmetry the mass splitting between $Z_c$ and $W_c$ should be the same as between $J/\psi$ and $\eta_c$,
so that the expected mass of this lower resonance is
\be
M(W_c) \approx 3785\,{\rm MeV}~,
\label{mwc}
\ee
and the expected dominant decay is $W_c \to \pi \eta_c$ with the same rate as $Z_c(3900) \to \pi J/\psi$. It should be noted that $W_c$ has to have the $G$-parity opposite to that of $Z_c$, so that it cannot be produced in association with a pion in decays of $Y(4260)$. It can however be produced from higher $1^{--}$ charmonium-like states in association with a $\rho$ meson, e.g. $e^+e^- \to \rho^\pm W_c^\mp$. 

Assuming the described simplest picture of the embedding for hadro-quarkonium, the quantum numbers of the $W_c$ should be $I^G(J^P)=1^-(0^+)$. A state with such quantum numbers and with mass given by Eq.(\ref{mwc}) is certainly prone to a strong decay into $D \bar D$ pairs. However, in the hadro-charmonium picture the decay into open charm channels is expected to be inhibited~\cite{dgv} by the suppressed probability of the reconnection of the bindings between heavy and light quarks. It can therefore be expected that the total width of the $W_c$ is not excessive as to prevent its observation in future experiments.

It can be also noted that a  resonance related to $Z_c(3900)$ with a lower mass is also expected in the tetraquark scheme~\cite{fmpppr} and for a $J^P=1^+$ molecule~\cite{mvwb}. In the former scheme the expected~\cite{fmpppr} mass ``is about 100\,MeV below '' the $Z_c$ resonance, which puts it distinctively higher in mass than the hadro-charmonium prediction (\ref{mwc}). In the molecular scheme the $J^P=0^+$ molecule should be above the $D \bar D$ threshold by approximately the same amount as the $J^P=1^+$ one is above the $D^* \bar D$ threshold, which puts it at approximately 3760\,MeV, i.e. distinctively lower than given by Eq.(\ref{mwc}). It thus can be expected that a search for an $I^G=1^-$ charmonium-like resonance in the mass range $3750 \div 3810\,$MeV will be helpful in resolving between the models of four-quark resonances.

The main conclusion from the discussion presented in this paper is that a further experimental study of the $Z_c(3900)$ resonance and related processes is vitally important for building an understanding of dynamics of multiquark heavy-light systems. It is argued that the most interesting aspects of such studies at the c.m. energy corresponding to $Y(4260)$ are
\begin{itemize}
\item{establishing the spin and parity of $Z_c(3900)$;}
\item{a search for a peak around 4030\,MeV in the $\pi J/\psi$ invariant mass spectrum in the process $Y(4260) \to \pi \pi J/\psi$;}
\item{a measurement of the branching fraction for decays of $Z_c(3900)$ into heavy meson pairs, $Z_c \to D^{*+} \bar D^0, D^+ \bar D^{*0}$;}
\item{a measurement of the rate of the decay $Z_c(3900) \to \pi \psi'$ relative to that of $Z_c(3900) \to \pi J/\psi$;}
\item{a search for the decays $Z_c(3900) \to \pi h_c$ and $Z_c \to \rho \eta_c$.}
\end{itemize}
Additionally at a higher energy of the $e^+e^-$ beams a search for the hypothetical  $W_c$ resonance with the mass in the range $3750 \div 3810\,$MeV can be performed  using the process $e^+e^- \to \rho W_c \to \rho \pi \eta_c$. As discussed above,  quantitative data from these studies would allow to resolve between the models of the internal structure of the $Z_c(3900)$ resonance.

I thank Alexander Bondar, Dan Cronin-Hennessy and Ron Poling for helpful discussions.
This work is supported, in part, by the DOE grant DE-FG02-94ER40823.

\end{document}